# Interstellar Low Energy Antiprotons as a Probe of Dark Matter and Primordial Black Holes

**Jonathan F. Ormes[1], Alexander Moiseev[1,2] and James Wells[3,4]**
[1]*NASA/Goddard Space Flight Center, Greenbelt, MD 20771, USA*
[2]*University Space Research Association, Seabrook, MD 20706, USA*
[3]*CERN, Theory Division, CH-1211, Geneva 23, Switzerland*
[4]*University of California, Davis, CA 95616-8677, USA*

**Abstract**
Cosmic ray antiprotons can originate from dark matter annihilating into quarks that subsequently decay into antiprotons. Evaporation of primordial black holes also can produce a significant antiproton flux. Since the spectrum of secondary antiprotons from cosmic ray interactions peaks at ~ 2 GeV and goes down sharply at lower energy, there is a window at energies < 1 GeV in which to look for excess antiprotons as a signature of these exotic antiproton sources. However, in the vicinity of the Earth low energy particles are strongly modulated by the solar wind, which makes any analysis ambiguous. The adverse effects of the solar wind can be avoided by placing a low energy antiproton spectrometer aboard an interstellar probe. The theoretical predictions are reviewed and the preliminary design of a light-weight, low-power instrument to make the measurements and a summary of the anticipated results are given in this paper.

## 1. Introduction.

The discrepancy between the observed large scale galaxy dynamics and theoretical modeling have converged to the problem of "missing mass" in the Universe (Trimble, 1987; Sikivie, 1995). This "missing mass", or dark matter, could be a significant fraction of the total Universe mass. Measurements of galactic rotation curves show that there is also "missing mass" on a galactic scale. One possible form of dark matter could be Weakly Interacting Massive Particles (WIMP). One way to make a direct detection of a WIMP would be to detect the recoil momentum from collisions in a cryogenic detector (Cabrera, 1998). Another promising method is to detect energetic neutrinos from WIMP annihilation. This dark matter can annihilate and decay in the galactic halo yielding antiprotons, positrons and photons. Here we will discuss what can be learned about galactic dark matter from the antiproton measurements and how this experiment could be carried out (Wells, Moiseev & Ormes, 1999).

## 2. Antiproton flux

The first experimental evidence for cosmic ray antiprotons is dated in the late 1970's when two independent experiments claimed detections using balloon-borne experiments (Golden et al., 1979; Bogomolov et al., 1979). At that time there was speculation that antiprotons had a primary origin in an anti-world. Antiprotons in the cosmic radiation could be from domains of antimatter, be from WIMPs or PBH, or be produced in cosmic ray interactions. People were excited by the implications of antiprotons, and considerable effort has since been made to clarify their origin. The early experiments were lacking in convincing particle identification and/or statistics (Buffington et al., 1981; Mitchell et al., 1995; Salamon et al., 1990; Streitmatter et al., 1989). The measurements were compared to calculations of the collisionally produced antiprotons based on a number of assumptions and uncertainties (Gaisser & Shaeffer, 1992; Stephens & Golden, 1987; Webber & Potgieter, 1989). The primary cosmic ray flux, the density and distribution of interstellar matter, and the mean and distribution of paths taken by cosmic rays are not precisely known. No reliable conclusion could be made based on the comparison of those early experiments and calculations. The Japanese-US balloon-borne experiment BESS has revolutionized this study (Yoshimura et al., 1995; Moiseev et al., 1997; Matsunaga et al., 1998). About 1000 antiprotons have now reliably been detected in 5 balloon flights. The measured antiproton fluxes are in a very good agreement with modern calculations (Orito et al., 1999). The secondary origin of antiprotons is proven except for one "but". The spectrum is flatter below 1 GeV and there is an excess flux at the $3\sigma$ level over the predicted. At this energy the cross section for antiproton production in proton-nucleus interactions goes down sharply (kinematic threshold), so any excess of the flux at this energy could be a signature of some "anomalous" component of antiproton flux. A number of authors have calculated the antiproton flux produced by primordial black hole decay (Hawking, 1974; Kiraly et al., 1981; Turner, 1982; Maki, Mitsui & Orito, 1996; and others) or in WIMP annihilations (Stecker, Rudaz & Walsh, 1985; Rudaz & Stecker, 1988; Jungman & Kamionkowski, 1994; Bottino et al., 1998). All the calculated spectra are flat at low energy; the absolute scaling depends on a number of parameters. With some parameters the flux from dark matter annihilation/PBH

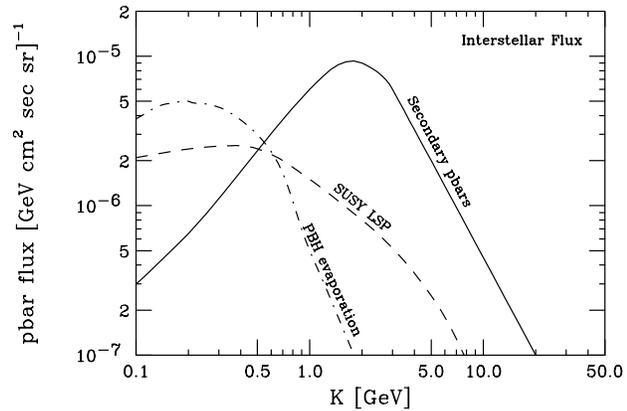

Fig.1. The antiproton spectrum (solid line – Simon, Molnar & Roesler 1998), from supersymmetric annihilations with $m_\chi$=62 GeV (dashed line – Bottino et al. 1998), and PBH evaporation (dash-dotted line – Maki, Mitsui & Orito 1996)

evaporation could dominate that from secondary production (fig.1). An alternative explanation for the flat antiproton spectrum is solar modulation, which produces a very similar shape (Labrador & Mewaldt, 1997; Bieber et al., 1999). Particles with different charge sign maybe modulated differently in a manner that depends on the orientation of the large scale solar magnetic field. Because of the scientific importance any statement concerning existence of "exotic sources", there must be no alternate explanations for the low energy antiproton excess flux.

## 3. Out-of-heliosphere experiment

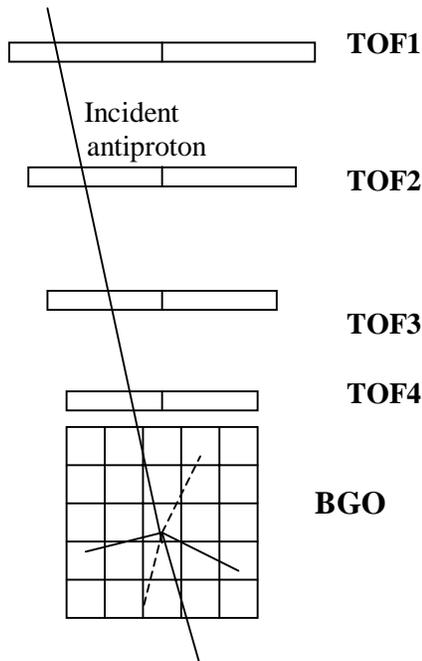

Fig.2. Scheme of proposed instrument

One way to avoid the influence of solar modulation is to measure the low energy antiproton flux outside the heliosphere. The planned interstellar probe could make such an experiment possible (Mewaldt et al., 1995). Interstellar probe is expected to leave the solar system at 20 AU/year and to reach a region free of the influence of the solar magnetic field within 5-10 years. Experiments on this probe will have to operate on very limited mass and electrical power resources but will collect data for many years so they can be made small.

The background requirements for an antiproton experiment in interstellar space are very difficult – the secondary antiproton/proton ratio at 0.2-0.5 GeV, where we expect to observe, is $\sim 10^{-6}$, so an instrument must be capable of separating protons from antiprotons with the power of at least $10^7$. The constraints of the very limited weight and power dictate the design. We propose to use the annihilation signature of < 0.2 GeV antiprotons that stop in a block of heavy scintillator (BGO) and release their entire rest mass energy (≈ 0.938 GeV). A cube of BGO with a side of 5 cm weighs 0.9 kg and stops antiprotons of energy < 0.2 GeV. A time-of-flight system (TOF) selects particles of these energies. The low energy limit is ~ 0.05 GeV set by the range an antiproton (proton) needs to pass through the TOF counters.

The separation of antiprotons from protons is key. Any low energy (< 0.2 GeV) proton which would pass TOF selections cannot deposit more than its own kinetic energy in the block; antiprotons will be required to deposit more than 0.3 GeV. So, to be accepted as an antiproton, the detected particle should be slow (E< 0.2 GeV) and deposit more than 0.3 GeV in the BGO calorimeter.

The proposed instrument is shown in fig.2. The choice of ~ 0.2 GeV as the highest antiproton energy to be detected is a compromise between energy band width, experimental mass and time-of-flight resolution. We have chosen dimensions of the crystal 5 cm × 5 cm × 5 cm. The calorimeter is subdivided into cubes 1 cm on a side to remove high-Z low energy nuclei which would deposit energy in a predictable, continuous pattern. The subdivision also removes protons which have energy a little above TOF threshold (0.2 GeV) but still pass TOF selections due to a measurement fluctuation, and deposit more than the calorimeter threshold energy (0.3 GeV) due to large scattering (or interaction) in the crystal. The TOF system consists of four 5 mm thick plastic scintillators spaced by 5 cm. The scintillator closest to the BGO crystal is 5cm×5cm and is divided into 2 strips, while the outer scintillator is larger, 8cm×8cm. The event trigger is the coincidence of all possible pair combinations (6) of TOF detectors and must be above a time-duration threshold which corresponds to β≃ 0.7. Moreover, all pulse heights from the 4 scintillators should lie in a band which corresponds to the ionization loss for the appropriate velocity particle. Finally, the trigger will require that the energy detected by the BGO crystal be above ~ 0.3 GeV, ~30% of the annihilation energy released. The resolution of the TOF is assumed to be 50 picosecond which is probably the best that can be presently achieved with scintillators of this size. All possible ways to reach and improve on this resolution should be explored. Currently we have simulation results that indicate a proton rejection power of $2\times10^6$ with good hope of reaching the requirement by tightening selections and finetuning the instrument design. It remains to study the possible background due to rare chance coincidences that could simulate an antiproton. The efficiency of the antiproton acceptance after all the selections are applied is shown in Fig.3. The energy resolution is ~10% at 0.1 GeV, provided by TOF. The estimated dimensions of the instrument are 20cm×15cm×10cm with <2kg weight. The goal is to use specialized electronics to keep the power consumption down. We estimate we can reach the 1.5-2 Watt range. The data rate would be ~ 300 bit/day.

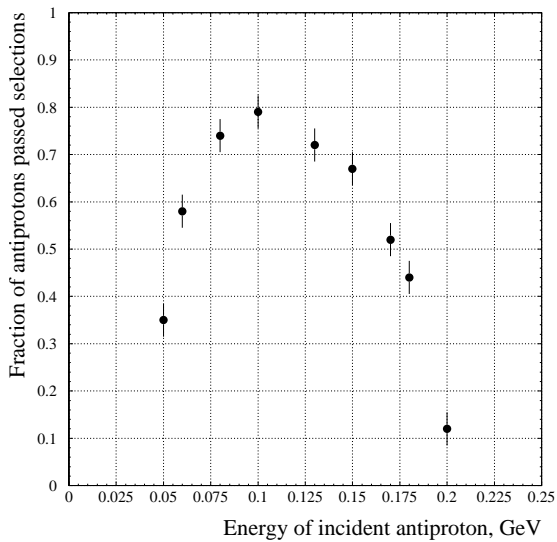

Fig.3. Efficiency of the antiproton acceptance

## 4. Expected results and conclusion

The geometrical factor of this instrument is ~ 3 cm²sr, and the expected event rate would be 0.1-1 antiprotons per day between 0.05 and 0.2 GeV. Approximately three years of observation would be required to obtain 10% statistical precision measurement of the flux of $5\times10^{-7}$ cm² s sr as shown in fig.1 for the PBH and WIMP signals. An observation of an excess of low energy antiprotons over and above that expected from the secondary production mechanism outside the heliosphere free of solar modulation effects would remove solar modulation as an explanation for the excess low energy antiprotons.